\documentclass[aps,pra,showpacs,amssymb,superscriptaddress,nofootinbib,twocolumn]{revtex4-1}

\usepackage{overpic,color}
\pdfoutput=1

\usepackage{amsmath,amsfonts,amsthm,amssymb}
\usepackage{braket}
\usepackage{setspace}
\usepackage{amsmath}
\usepackage{appendix}
\usepackage[utf8]{inputenc}
\usepackage{bbm}
\usepackage{bm}
\usepackage{amsbsy}
\usepackage{dsfont} 
\usepackage{graphicx} 
\usepackage{epsfig}
\usepackage{epstopdf}
\usepackage{dsfont}
\usepackage{color}
\usepackage[bookmarks=true, colorlinks]{hyperref}
\makeatletter
\newcommand\org@hypertarget{}
\let\org@hypertarget\hypertarget
\renewcommand\hypertarget[2]{%
  \Hy@raisedlink{\org@hypertarget{#1}{}}#2%
  }
\makeatother

\usepackage[figure,table]{hypcap}
\usepackage{enumerate}
\usepackage[resetlabels]{multibib}

\hypersetup{
	bookmarksnumbered,
	pdfstartview={FitH},
	citecolor={darkgreen},
	linkcolor={darkred},
	urlcolor={darkblue},
	pdfpagemode={UseOutlines}}
\definecolor{darkgreen}{RGB}{50,190,50}
\definecolor{darkblue}{RGB}{0,0,190}
\definecolor{darkred}{RGB}{238,0,0}
\usepackage{soul}

\begin{document}

\title{Light-matter entanglement over 50km of optical fibre}

%

\title{Light-matter entanglement over 50~km of optical fibre}

\vspace{20mm}


\author{V. Krutyanskiy$^\dagger$}

\affiliation{Institut f\"ur Quantenoptik und Quanteninformation,\\
	\"Osterreichische Akademie der Wissenschaften, Technikerstr. 21A, 6020 Innsbruck,
	Austria}

\author{M. Meraner$^\dagger$}

\affiliation{Institut f\"ur Quantenoptik und Quanteninformation,\\
	\"Osterreichische Akademie der Wissenschaften, Technikerstr. 21A, 6020 Innsbruck,
	Austria}
\affiliation{
	Institut f\"ur Experimentalphysik, Universit\"at Innsbruck,
	Technikerstr. 25, 6020 Innsbruck, Austria}

\author{J. Schupp$^\dagger$}

\affiliation{Institut f\"ur Quantenoptik und Quanteninformation,\\
	\"Osterreichische Akademie der Wissenschaften, Technikerstr. 21A, 6020 Innsbruck,
	Austria}
\affiliation{
	Institut f\"ur Experimentalphysik, Universit\"at Innsbruck,
	Technikerstr. 25, 6020 Innsbruck, Austria}

\author{\\V. Krcmarsky}

\affiliation{Institut f\"ur Quantenoptik und Quanteninformation,\\
	\"Osterreichische Akademie der Wissenschaften, Technikerstr. 21A, 6020 Innsbruck,
	Austria}
\affiliation{
	Institut f\"ur Experimentalphysik, Universit\"at Innsbruck,
	Technikerstr. 25, 6020 Innsbruck, Austria}

\author{H. Hainzer}

\affiliation{Institut f\"ur Quantenoptik und Quanteninformation,\\
	\"Osterreichische Akademie der Wissenschaften, Technikerstr. 21A, 6020 Innsbruck,
	Austria}
\affiliation{
	Institut f\"ur Experimentalphysik, Universit\"at Innsbruck,
	Technikerstr. 25, 6020 Innsbruck, Austria}

\author{B. P. Lanyon}
\email{ben.lanyon@uibk.ac.at,$\dagger$ These authors contributed equally}

\affiliation{Institut f\"ur Quantenoptik und Quanteninformation,\\
	\"Osterreichische Akademie der Wissenschaften, Technikerstr. 21A, 6020 Innsbruck,
	Austria}
\affiliation{
	Institut f\"ur Experimentalphysik, Universit\"at Innsbruck,
	Technikerstr. 25, 6020 Innsbruck, Austria}


\begin{abstract}
	\vspace{5mm}

\noindent When shared between remote locations, entanglement opens up fundamentally new capabilities for 
science and technology \cite{Kimble2008, Wehnereaam9288}. 
Envisioned quantum networks distribute entanglement between their remote matter-based quantum nodes, in which it is stored, processed and used \cite{Kimble2008}.  
Pioneering experiments have shown how photons can distribute entanglement between single ions or single atoms a few ten meters apart \cite{Moehring2007, Ritter2012} and between two nitrogen-vacancy centres 1 km apart \cite{Hensen2015}. 
Here we report on the observation of entanglement between matter (a trapped ion) and light (a photon) over 50~km of optical fibre: a practical distance to start building large-scale quantum networks. 
Our methods include an efficient source of light-matter entanglement via cavity-QED techniques and a quantum photon converter to the 1550~nm telecom C band. 
Our methods provide a direct path to entangling remote registers of quantum-logic capable trapped-ion qubits \cite{Duan2010, Sangouard2009, Friis2018}, and the optical atomic clock transitions that they contain \cite{PhysRevLett.102.023002, Chen2017}, spaced by hundreds of kilometers.
\end{abstract}
\maketitle

\begin{figure*}[t]
	\vspace{0mm}
	\begin{center}
		\includegraphics[width=1\textwidth]{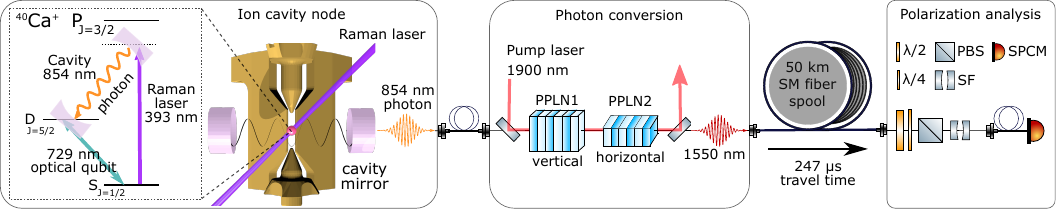}
		\vspace{0mm}
		\caption{
			\textbf{Simplified experiment schematic.} 
			From left to right: A single atomic ion (red sphere) in the centre of a radio-frequency linear Paul trap (gold electrodes) and a vacuum antinode of an optical cavity. A Raman laser pulse triggers emission of an 854~nm photon into the cavity, which exits to the right. The photon, polarisation-entangled with two electronic qubit states of the ion (two Zeeman states of the D$_{J=5/2}$ manifold, not shown), is then wavelength-converted to 1550~nm using difference frequency generation involving ridge-waveguide-integrated periodically-poled lithium niobate (PPLN) chips and a strong ($\sim$~1 W) pump laser at 1902~nm \cite{Krutyanskiy2017}. The photon then passes through a 50~km single-mode fibre spool, is filtered with a $250$~MHz bandwidth etalon (SF) to reduce noise from the conversion stage \cite{Krutyanskiy2017}, and is polarisation-analysed using waveplates, a polarising beam splitter (PBS) and a solid-state single photon counting module (SPCM, InGaAs \emph{ID230} from IDQuantique). The electronic state of the ion is measured (not shown), conditional on the detection of a photon. Additional photon conversion filters are not shown \cite{Krutyanskiy2017}.  For further details see \cite{SuppMat}.
		}
		\label{figure2}
		\vspace{-6mm}
	\end{center}
\end{figure*}

Our network node consists of a $^{40}$Ca$^+$ ion in a radio-frequency linear Paul trap with an optical cavity that enhances photon collection on the 854~nm electronic  dipole transition. (Figure ~\ref{figure2}).
A Raman laser pulse at 393~nm triggers emission, by the ion, of a photon into the cavity via a bichromatic cavity-mediated Raman transition (CMRT) \cite{Stute2012}.
Two indistinguishable processes are driven in the CMRT, each leading to the generation of a cavity photon and resulting in entanglement between photon polarisation and the electronic qubit state of the ion of the form $1/\sqrt{2}~(\ket{D_{J{=}5/2,~m_j{=}-5/2},V}+\ket{D_{J{=}5/2,~m_j{=}-3/2},H})$, with horizontal ($H$) and vertical ($V$) photon polarisation and two metastable Zeeman states of the ion ($D_{J,~m_j}$) \cite{SuppMat}. 
The total probability of obtaining an on-demand free-space photon out of the ion vacuum chamber (entangled with the ion)  is 0.5 $\pm{0.1}$ \cite{SuppMat}. 

While the $\sim$ 3~dB/km losses suffered by 854~nm photons through state-of-the-art optical fibre allows for few km internode distances, transmission over 50~km would be $10^{-15}$.
854~nm photons are also frequency-incompatible with other examples of quantum matter, preventing the realisation of ion-hybrid quantum systems over any distance.
Single photon frequency conversion to the telecom C band (1550~nm) offers a powerful general solution: this wavelength suffers the minimum fibre transmission losses  ($\sim$ 0.18~dB/km) and is therefore an ideal choice for a standard interfacing wavelength for quantum networking. 
Photons from solid-state memories \cite{Maring2017}, cold gas memories \cite{Radnaev:2010br, Albrecht2014}, quantum dots and nitrogen-vacancy centres \cite{PhysRevApplied.9.064031} have been converted to telecom wavelengths.
Frequency conversion of photons from ions has recently been performed, including to the telecom C band (without entanglement) \cite{Walker2018}, to the telecom O band with entanglement over 80~m \cite{Bock2018} and directly to an atomic Rubidium line at 780~nm \cite{Siverns2018}.

We inject single-mode fibre-coupled photons from the ion into a polarisation-preserving photon conversion system (previously characterised using classical light \cite{Krutyanskiy2017}).
In summary, a $\chi_2$ optical nonlinearity is used to realise difference frequency generation, whereby the energy of the 854~nm photon is reduced by that of a pump-laser photon at 1902~nm, yielding 1550~nm. 
Two commercially-available free-space and crossed PPLN ridge waveguide crystals are used, one to convert each polarisation, in a self-stable polarisation interferometer. 
The total fibre-coupled device conversion efficiency here is 25~\% for an added white photon noise of 40~cps, within the  filtering bandwidth of 250~MHz centred at 1550~nm.

Following conversion, the telecom photon is injected into a 50.47~km `SMF28' single-mode fibre spool with 0.181~dB/km loss
($10.4\pm 0.5$\% measured total transmission probability). 
The spool is not actively stabilised. 
Polarisation dynamics in an unspooled fibre could be actively controlled using methods developed in the field of quantum cryptography (e.g. \cite{Treiber09}). 
Finally, free space projective polarisation analysis is performed and the photon is detected using a telecom solid-state photon detector with an efficiency of 0.10$\pm{0.01}$ and free-running dark count rate of ~2 cps. 
Measurement of the ion-qubit state is performed conditional on the detection of a 50~km photon within a 30~$\mu$s time window: the Zeeman ion qubit is mapped into the established $^{40}$Ca$^+$ optical clock qubit \cite{Schindler2013} via laser pulses at 729~nm, followed by standard fluorescence state detection \cite{SuppMat}. %
%

Quantum state tomography is performed to reconstruct the two-qubit (ion qubit and photon polarisation qubit) state \cite{SuppMat}. 
The 247~$\mu$s photon travel time through the fibre limits the maximum attempt rate for generating a photon from the ion to 4~kHz (2~kHz if the fibre was stretched out away from our ion to force an additional delay for the classical signal `photon click' to return). 
Here, until photon detection occurs, photon generation is (Raman laser pulses are) performed every 453 $\mu$s, yielding an attempt rate of 2.2 kHz. 
For the complete experimental sequence see \cite{SuppMat}.
All error bars on quantities derived from the tomographically-reconstructed states (density matrices) are based on simulated uncertainties due to finite measurement statistics \cite{SuppMat}.

\begin{figure}[t]
	\vspace{0mm}
	\begin{center}
		\includegraphics[width=\columnwidth]{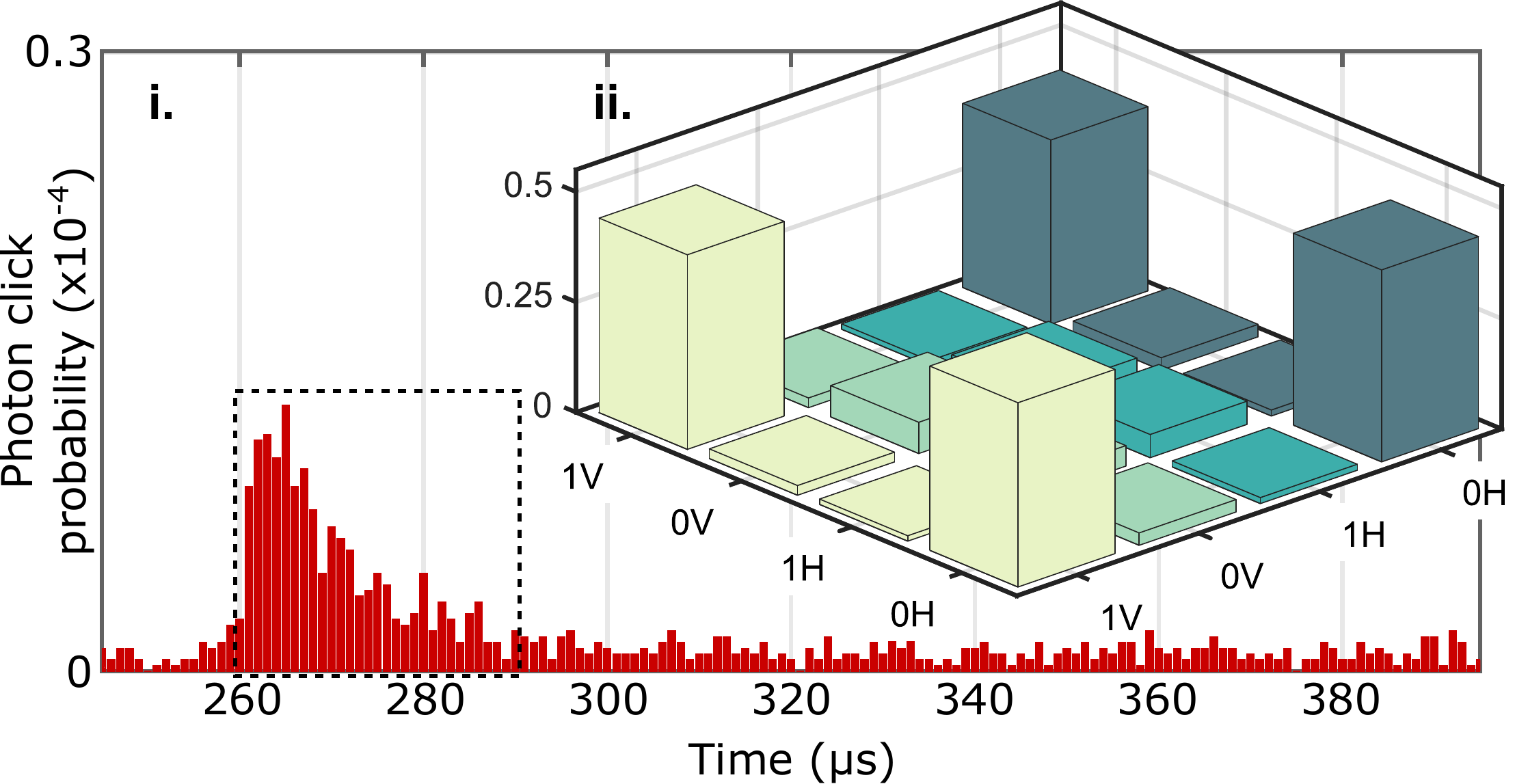}
		\vspace{-5mm}
		\caption{\textbf{Observation of ion-photon entanglement over 50~km of optical fibre}. 
			\textbf{i.} 2D red bar chart: histogram of photon detection times (photon wavepacket in dashed box), following the generation of an 854~nm photon with a 30~$\mu$s Raman laser pulse (R) $\approx$ 250 $\mu$s earlier, repeated at 2.2~kHz. ion-photon state tomography is performed for photon detection events recorded in the dashed box (total contained probability $P=5.3 \times 10^{-4}$).  
\textbf{ii}. 3D bar chart: absolute value of experimentally-reconstructed density matrix of the telecom photonic polarisation qubit ($H$ and $V$ are Horizontal and Vertical, respectively) and ion-qubit state ($\ket{0}=\ket{D_{J=5/2,m_j=-3/2}}$, $\ket{1}=\ket{D_{J=5/2,m_j=-5/2}}$).
		}
		\label{figure3}
		\vspace{-5mm}
	\end{center}
\end{figure}

A strongly entangled ion-photon state is observed (Figure \ref{figure3}) over 50~km, quantified by a concurrence \cite{PhysRevLett.80.2245} $C{=}0.75 \pm{0.05}$ and 
state fidelity $F^{m}{=}0.86 \pm{0.03}$ with a maximally entangled state ($C{=}1$).  
Simulating a CHSH Bell inequality test \cite{PhysRevLett69_CHSH} on our tomographic data yields a value of 2.43$\pm{0.127}$, thereby exceeding the classical bound (of 2) by 2.4 standard deviations. Using a shorter detection window (first 2/3 of the full photon wavepacket) increases the signal to noise ratio and yields $F^{m}{=}0.90\pm 0.03$ and CHSH Bell inequality violation by 4.8 standard deviations at the expense of an efficiency decrease of only 10\%.
The quality of our light-matter entangled state passes the most stringent thresholds for its subsequent application.

For a detailed analysis of the sources of infidelity in the entangled state see \cite{SuppMat}; here now is a short summary.
In a second experiment, the telecom entangled stated is reconstructed right after the conversion stage (without the 50~km spool), yielding $F^{m}{=}0.92 \pm{0.02}$. 
The drop in fidelity when adding the 50~km spool can, to within statistical uncertainty, be entirely explained by our telecom photon detector dark counts (2~cps). 
In a third experiment, the 854~nm entangled state is reconstructed right out of the vacuum chamber (without conversion), yielding $F^{m}{=}0.967 \pm{0.006}$.
The observed drop in fidelity through the conversion stage alone is dominated by added photon noise (caused by Anti-Stokes Raman scattering of the pump laser \cite{Krutyanskiy2017}). 
The infidelity in the 854~nm state is consistent with that achieved in \cite{Stute2012}. 

The total probability that a Raman pulse led to the detection of a photon after 50~km was $P=5.3 \times 10^{-4}$, which given an attempt rate of 2.2~kHz yielded a click rate of $\approx$ 1 cps. 
Photon loss mechanisms in our experiment are discussed in \cite{SuppMat}. 
In summary, the 50~km fibre transmission (0.1) and our current telecom detector efficiency (0.1) limit the maximum click probability to $P=0.01$.
The majority of other losses are in passive optical elements, and could largely be eliminated by e.g. more careful attention to coupling into optical fibres and photon conversion waveguides. 
In combination with state-of-the-art telecom detectors (efficiency ~0.9 for $<$~5 dark cps), a total 50~km efficiency of $P\approx$ 0.01 would be expected and a corresponding click rate of $\approx$ 20~cps. 

\begin{figure}[ht]
	\vspace{7mm}
	\begin{center}
		\includegraphics[width=\columnwidth]{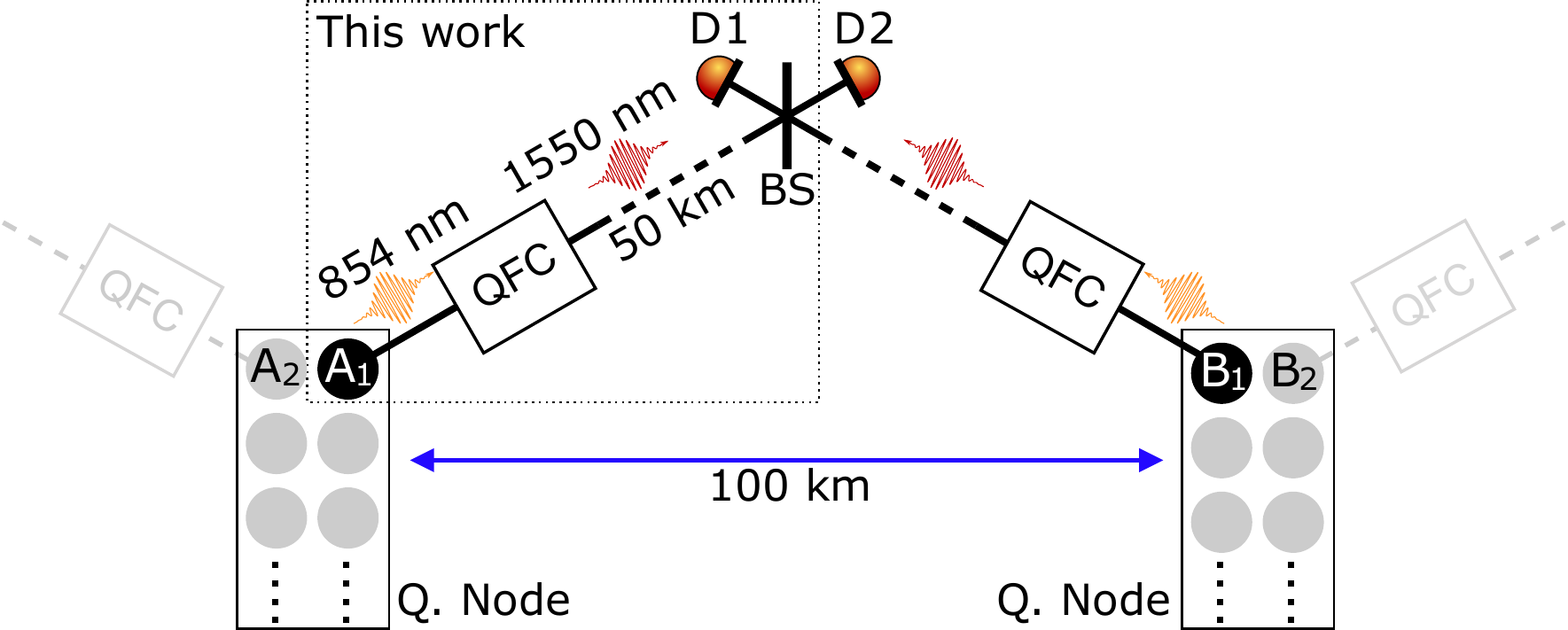}
		\vspace{-2mm}
		\caption{
			\textbf{Path to 100~km matter-matter entanglement.}    
			This work: quantum frequency conversion (QFC) converts a photon, emitted on-demand from and entangled with an ion qubit (A$_1$) in node A, to the telecom C band at 1550~nm. The photon then travels through 50~km of optical fibre before detection (D1 or D2). Future work: duplicating the system, interfering the two photonic channels on a beamsplitter (BS). Single or two photon detection heralds the projection of ions A$_1$ \& B$_1$ into an entangled state \cite{Luo2009}. Deterministic intra-node quantum logic and measurement between e.g. B$_1$ \& B$_2$ and A$_1$ \& A$_2$ can swap the entanglement over larger distances (quantum repeater). Additional qubits in nodes are available for entanglement purification. Nodes could as well contain solid-state memories \cite{Seri2017}, NV centres \cite{PhysRevApplied.9.064031} or neutral atoms \cite{Radnaev:2010br, Albrecht2014}. 
		}
		\label{figure1}
		\vspace{-6mm}
	\end{center}
\end{figure}

The rates for future 100~km-spaced photon-detection heralded ion-ion entanglement using our methods are now discussed (see Figure \ref{figure1}).
A modestly optimised version of our system is considered, that achieves an on-demand 50~km photon click probability of $P=0.01$ and operates at an attempt rate of $R=2$~kHz (the two-way light travel time). 
By duplicating our experiment, and following a two-photon click heralding scheme \cite{Luo2009}, the probability of heralding a 100~km spaced ion-ion entangled state would be $H_2=\frac{1}{2}P^2= 5 \times 10^{-5}$, at an average click rate of $H_2\times R = 0.1$~cps (comparable with the first rates achieved over a few meters  \cite{Matsukevich2008} of 0.03 cps). 
Following a single-photon click scheme \cite{Luo2009}, one finds $H_1=2P\times0.1=0.002$, and an average click rate of 2~cps, where 0.1 is the reduced photon generation probability at each node (as required for this scheme). 
This factor 20 improvement over the two-photon scheme comes at the expense of the need to interferometrically stabilise the optical path length across the 100~km network. 
The threshold value $P>0.04$ for which $H_2>H_1$ is within reach with our setup, when allowing for recently-developed 0.16 dB/km loss telecom C band fibres. 

An approach to significantly increase the remote entanglement heralding rate is multi-mode quantum networking, where many photons are sent, each entangled with different matter qubits. In this way, of running many such processes in parallel, the probability of at least one successful heralding event occurring can be made arbitrarily high.  
In our setup, for example, multiple ions can be trapped and it may be possible to produce a train of photons, each entangled with a different ion. 
In this case, a higher rate of photon production can be employed, as the time between photons in the train is not limited by the light travel time. 
Furthermore, multi-mode networking could be realised using inhomogenously-broadened ensemble based solid state quantum memories \cite{PhysRevLett.98.190503}. Such memories could be quantum-networked with ions via a photon conversion interface \cite{Seri2017} to form a powerful hybrid system for long distance quantum networking.

Decoherence of the ion-qubit plays no significant role during the 250 $\mu$s (`storage time') required for the 50~km distribution. 
Establishing entanglement between any two nodes of a quantum network requires storage times in each node at least equal to the light travel time between them. 
Additional measurements demonstrate ion-qubit storages times of at least 20~ms in our system \cite{SuppMat}, opening up the possibility of quantum networks several thousands of kilometers in extent.

The 50~km photon in our experiments is entangled with the 729~nm optical qubit clock transition in $^{40}$Ca$^+$, over which a fractional frequency uncertainty of $1 \times 10 ^{-15}$ has been achieved (comparable with the Cs standard) \cite{PhysRevLett.102.023002}. Furthermore, $^{40}$Ca$^+$ can be co-trapped with Al$^+$ \cite{Guggemos2015}, which contains a clock transition for which a fractional systematic frequency uncertainty at the $1 \times 10 ^{-18}$ level was recently achieved  \cite{HumeICAP, Chen2017}.  Transfer of the remote $^{40}$Ca$^+$ entanglement to co-trapped Al$^+$ ion could be done via quantum logic techniques \cite{Chen2017, Schmidt749}. 
As such, our work provides a direct path to realise entangled networks of state-of-the-art atomic clocks over large distances \cite{Komar2014}. Entangling clocks provides a way to perform more sensitive measurements of their average ticking frequencies \cite{Komar2014} and to overcome current limits to their synchronisation \cite{Ilo-Okeke2018}.

\begin{acknowledgments}
	We thank the staff at IQOQI Innsbruck; Rainer Blatt for providing encouragement, laboratory space, and the environment and group support in which to develop our work; Daniel Heinrich, Klemens Sch\"uppert, Tiffany Brydges, Christine Maier and Tracy Northup for their support. 
	This work was supported by the START prize of the Austrian FWF project Y 849-N20, the Army Research Laboratory Center for Distributed Quantum Information via the project SciNet, the Institute for Quantum Optics and Quantum Information (IQOQI) of the Austrian Academy Of Sciences (OEAW) and the European Union's Horizon 2020 research and innovation programme under grant agreement No 820445 and project name `Quantum Internet Alliance'. The European Commission is not responsible for any use that may be made of the information this paper contains.

\end{acknowledgments}

\section*{Author Contributions}

All authors contributed to the design, development and characterisation of the experimental systems. In particular, JS focused on the ion trap and optical cavity, MM on the photon conversion system, VKrc on the ion trap, HH on laser frequency stabilisation and VKru and BPL on all aspects. 
Experimental data taking was done by VKru, VKrc, MM and JS. 
Data analysis and interpretation was done by VKru, JS, MM and BPL.
All authors contributed to the paper writing. 
The project was conceived and supervised by BPL.

\bibliography{references_50km.bib}


\hypertarget{sec:appendix}
\appendix
\section*{Supplementary Material}
\renewcommand{\thesubsubsection}{\Roman{subsection}.\Alph{subsubsection}}
\renewcommand{\thesubsection}{\Roman{subsection}}
\renewcommand{\thesection}{}
\setcounter{equation}{0}
\numberwithin{equation}{section}
\renewcommand{\thefigure}{A.\arabic{figure}}

\subsection{Experimental details}


\subsubsection{Ion trap}
\
We use a 3D radio-frequency linear Paul trap with a DC endcap to ion separation of $2.5~\rm{mm}$ and ion to blade distance of $0.8~\rm{mm}$.
The trap electrodes are made of titanium, coated with gold and are mounted on Sapphire holders. 
The trap drive frequency is $23.4~\rm{MHz}$.
The radial secular frequencies are  $\omega_{x} \approx \omega_{y}= 2\pi \times2.0~\rm{MHz}$ and split by approximately $10 ~\rm{kHz}$ and the axial frequency is $\omega_z=2\pi \times 0.927~\rm{MHz}$.

Atoms are loaded from a resistively heated atomic oven and ionised via a two photon process involving 375~nm and 422~nm laser light. 



\subsubsection{Cavity parameters}
\label{cavityparams}

The optical cavity around the ion is near-concentric with a length $l = 19.9057 \pm 0.0003$ mm and radii of curvature $ROC = 9.9841 \pm 0.0007$ mm, determined from simultaneous measurements of the free spectral range (FSR) and higher-order TEM mode spacing (assuming identical mirror geometries) \cite{siegman1986lasers}.
From this we calculate an expected cavity waist of $\omega_0 = 12.31 \pm 0.07$ $\mu$m and a maximum ion-cavity coupling rate of $g_{max} = 2\pi \cdot 1.53 \pm 0.01$ MHz. 
At a wavelength of 854 nm, the finesse of the TEM$_{00}$ mode is $\mathcal{F} = \frac{2\pi}{\mathcal{L}} = 54000 \pm 1000$, with the total cavity losses $\mathcal{L} = T_1 + T_2 + l_{1+2} = 116 \pm 2 $ ppm, determined from measurements of the cavity ringdown time $\tau_C = \frac {\mathcal{F}}{\pi}\cdot \frac l{c_0}$, with $c_0$ the speed of light in vacuum. From this one can calculate the cavity linewidth $2\kappa = 2\pi \cdot 140 \pm 3$ kHz, $\kappa$ being the half-width at half maximum.

Comparison of the spontaneous scattering rate of the $P_{3/2}$ state of the ion ($\gamma=2\pi\cdot 11.45$ MHz, half width) with $g_{max}$ shows that the system is far away from the strong coupling regime. However, the cooperativity $C=\frac{g_{max}^2}{2\kappa \gamma} = 1.47\pm 0.03$. 

The transmission $T_{1,2}$ of our cavity mirrors\footnote{Polishing of the mirror substrates done by Perkins Precision Development, Boulder (Colorado). Coating done by Advanced Thing Films.} was verified by applying the method described in \cite{Hood}, yielding $T_1 = 2.2 \pm 0.3 $ ppm, $T_2 = 97 \pm 4 $ ppm, such that the combined mirror losses from scattering and absorption $l_{1+2} = 17 \pm 5$ ppm. 

From this one can calculate the probability that a photon inside the cavity exits through mirror $T_2$ (designated output mirror) as $P_{out}^{max}=T_2/(T_1+T_2+L_{1+2})=0.83 \pm 0.03$.  $P_{out}^{max}$ is therefore the maximum photon collection probability from the ion in our system (with the current mirrors). 

The cavity length is stabilised via the Pound-Drever-Hall (PDH) method \cite{Black} to a laser at 806 nm with a linewidth on the order of $1~$kHz \cite{HeleneMasterThesis}.
The 806~nm wavelength lies far from any transition in $^{40}$Ca$^{+}$ to minimize AC Stark-shifts on the ionic transitions. The cavity is locked to a TEM$_{01}$ mode and the ions sits in the central intensity minimum to further minimise AC stark shifts. 

The cavity waist is centred on the ion via course tuning of a 3D piezo stick-slip translation-stage system (Attocube). 
Before experiments, photon generation efficiency is optimised by placing the ion in a cavity anti-node.
This is done via fine tuning of the cavity position along its axis by applying a small bias voltage to the corresponding piezo stage.


\subsubsection{Experimental geometry}
\label{Exp_geom}

The optical cavity axis lies perpendicular to the principle ion trap axis. A magnetic field of 4.22 G is applied perpendicular to the cavity axis and at an angle of $45$ degrees to the principle ion trap axis. The Raman photon generation beam is circularly polarised and parallel to the magnetic field (to maximise the coupling on the relevant dipole transition $S_{J=1/2,m_j=-1/2}\leftrightarrow P_{J=3/2,m_j=-3/2}$, see Figure \ref{levelscheme}).


\subsubsection{Pulse sequence for 50~km experiment}
\label{pulse_sequence}
\begin{figure*}[th]
	\centering
	\includegraphics[width=0.95\textwidth]{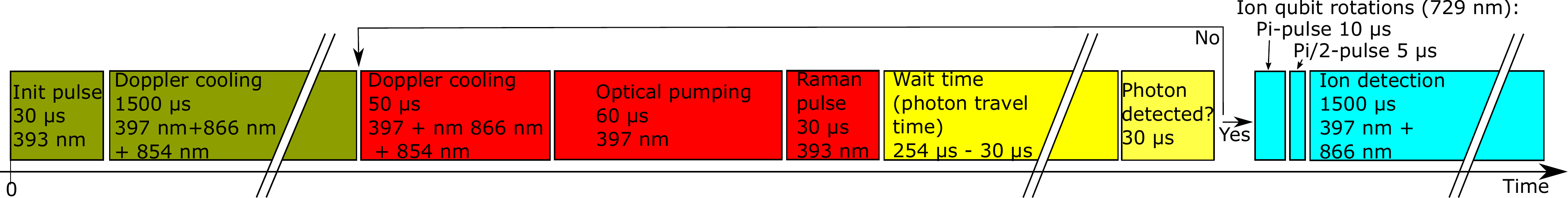}
	\caption{\textbf{Pulse sequence for the $50 ~\rm{k m}$ entanglement experiment.} 
The starting sequence (shown in green) consists of an initialisation laser pulse for intensity stabilisation and Doppler cooling. 
The loop (shown in red and yellow) consists of additional Doppler cooling, optical pumping, a (photon generation) Raman pulse and the conditional logic (waiting as the photon travels through the 50~km fiber to the detector and a detection window). 
If there is a photon detected within this window, ion qubit manipulation and state detection (shown in blue) are performed, otherwise the loop repeats.
} 
	\label{Pulse_sequnce_854}
\end{figure*}

\begin{figure}[th]
	\centering
	\includegraphics[width=\columnwidth]{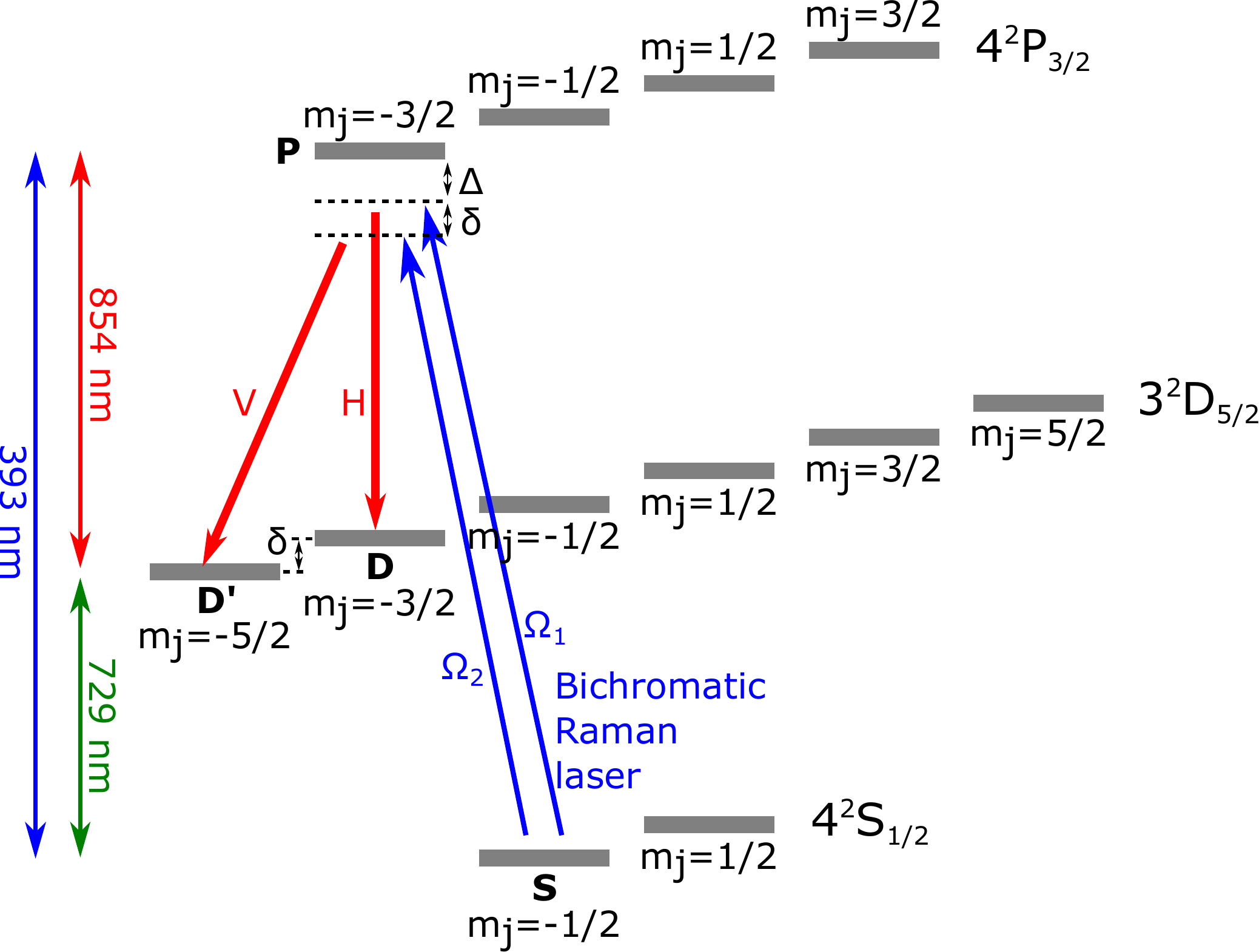}
	\label{fig:level_scheme}
	\caption{Relevant energy level scheme of $^{40}$Ca$^+$.  The cavity-mediated Raman transition (CMRT) is shown, for generating an 854~nm photon that is polarisation entangled with the final electronic state of the ion. For details on the CMRT see the publication \cite{Stute2012} and the PhD thesis \cite{stutethesis}. In summary, following optical pumping, the ion (single outer valence electron) begins in the state S.  A bichromatic 393~nm Raman pulse is applied with a frequency splitting equal to that of the D-D' states. 
The bichromatic field and an optical cavity locked to the 854~nm transition generate two Raman processes, leading to the total transformation $\ket{S,0} \rightarrow 1\sqrt{2}(\ket{D, H}+\ket{D',V}$, where $0$, $H$ and $V$ are: 0 photons, a single horizontally-polarised and a single vertically-polarised photon in the cavity, respectively.  The aforementioned polarisations are the projections into the optical cavity axis (perpendicular to the magnetic field). The relative amplitude of the two terms in the entangled superposition state are balanced in a separate calibration stage (see \cite{stutethesis}), controlled via the relative intensities of the two frequency components in the Raman beam. The phase of the entangled state can be controlled via the relative phase of the two frequency components in the Raman beam. 
The detuning $\Delta=409\pm 10$ MHz. 
In experiments we set the Rabi frequencies $\Omega_1$ and $\Omega_2$ so as to produce both polarisations with equal probabilities, as described in \cite{Stute2012}. The  total AC Stark shift exerted by the bichromatic Raman laser on the $S$ state was measured (via 729~nm spectroscopy) to be $AC=2\pi \cdot (1.14\pm 0.05)$ MHz, where $AC=\Omega_1^2/4\Delta+\Omega_2^2/4(\Delta+\delta)$. 
}
	\label{levelscheme}
\end{figure}

Fig. \ref{Pulse_sequnce_854} shows the laser pulse sequence for the 50~km ion-photon entanglement experiment. First, a $30~ \rm{\mu s}$ `initialisation' laser pulse at 393 nm is applied, measured by a photodiode in transmission of the ion-trap chamber, which allows for intensity stabilisation of the subsequent 393~nm photon generation Raman pulse via a sample and hold system. The initialisation pulse is followed by a $1500~ \rm{\mu s}$ Doppler cooling pulse, involving three laser fields as indicated.

Next, a loop starts in which single photons are generated. 
This loop consists of an additional Doppler cooling pulse ($50 ~\rm{\mu s}$), optical pumping to the $S=S_{J=1/2,m_j=-1/2}$ (see Fig \ref{Pulse_sequnce_854}) state via circularly polarised 397~nm laser light ($60 ~\rm{\mu s}$), and a $393 ~\rm{n m}$ photon generation Raman pulse ($30 ~\rm{\mu s}$). This is followed by a wait time for the photon to travel through the 50~km fibre and a subsequent photon detection window. 
This sequence loops until a photon is detected. 

In the case of a photon detection (detector `click'), the state of the ion is measured. 
To perform an ion state measurement, the D' electron population is first mapped to the S state via a 729 nm $\pi$ pulse (Fig. \ref{Pulse_sequnce_854}). That is, the D-manifold qubit is mapped into an optical qubit (with logical states S and D). In order to measure which of these states the electron is in, the standard electron shelving technique is used. That is, 397~nm laser light is sent to the ion (with an additional 866~nm re-pumper). If the electron is in S then it will scatter 397~nm photons, which are partially collected via a photo multiplier tube and one observes a bright ion. If the electron is in D (lifetime $\approx$ 1 s) then photons are scattered with negligible probability by the ion. We perform this measurement for a `detection time' (397~nm photon collection time) of $1500~\rm{\mu s}$, which is sufficient to distinguish bright (scattering) and dark (non-scattering) ions with an error less than 1\%. The aforementioned process implements a projective measurement into the eigenstates of the $\sigma_z$ basis (Pauli spin-1/2 operator). 

To perform measurements in other bases  e.g $\sigma_x$ ($\sigma_y$), as required for full quantum state tomography, an additional $\pi/2$ pulse on the S\textsubscript{m\textsubscript{j}=-1/2} to D\textsubscript{m\textsubscript{j}=-3/2} with a 0 ($\pi/2$) phase is applied after the $\pi$ pulse and before the 397~nm pulse, to rotate the qubit measurement basis.

\subsection{Photon distribution efficiency}
In this section information is presented on the efficiency with which photons are distributed in the 50~km experiment and the sources of photon loss.  
\subsubsection{Current setup efficiency}
\label{sec_current_efficiency}
In the 50~km experiment, the total probability that a Raman photon generation pulse leads to a photon click after 50~km is $5.3\times10^{-4}$ (After summing up the outcomes of all polarisation projections). 

The total probability of obtaining an on-demand free-space photon out of the ion vacuum chamber is $P_{out}=0.5 \pm$ 0.1.
This value is inferred from the measured efficiency with which we detect single-mode fibre-coupled (ion-entangled) photons at 854 nm (before the conversion stage), after correcting for the measured 1st fibre-coupling stage efficiency and the known 854~nm photon detector efficiency. 
The uncertainty in $P_{out}$ is dominated by the uncertainty in the 1st fibre-coupling stage efficiency, which could be reduced in future. 

The overall efficiency of the frequency-conversion setup, including spectral filtering, is $0.25 \pm 0.02$, measured with classical 854~nm light. 
For a detailed description see \cite{Krutyanskiy2017}. A short overview of the contributing photon losses are summarized in table \ref{losses}.
Multiplying all the transmissions together leads to a total expected probability of detecting the photon after 50~km of $(6.5 \pm 1.5)\times 10^{-4} $, which is consistent to within one standard deviation with the measured value of $5.3\times10^{-4}$.

\begin{table}[!h]
	\begin{tabular}{|c|c|c|}
		\hline
		Location in the photon path &Efficiency\\
		\hline On demand photon out of cavity $P_{out}$ & $0.5 \pm 0.1$\\
		1$^\text{st}$ single-mode fibre coupling & $0.5 \pm 0.1$\\
		Telecom conversion stage (\& filtering) & $0.25 \pm 0.02$\\
		50~km fibre transmission & $0.104 \pm 0.005$ \\
		Telecom photon detector efficiency & $0.10 \pm 0.01$\\
		\hline Expected 50~km detection probability & $(6.5 \pm 1.6) \times 10^{-4} $\\
		\hline
	\end{tabular}
	\label{losses}
	\caption{Photon losses in our 50~km photon distribution experiment. See Fig. 1 in the main paper for the respective locations in the experimental setup.} 
\end{table}

A total 50~km detection probability of 0.01 should be straightforward to achieve. For example, telecom photon detectors with efficiencies of $>$ 0.8 and dark count rates of $<$ 5 cps are now available commercially. Since taking the data presented in this paper, we have improved the 1st fibre-coupling stage efficiency to 0.9$\pm{0.1}$ and further improvements should be possible. These changes alone are sufficient to achieve a total 50~km efficiency above 0.01.  

The efficiency $P_{out}$ in our setup is limited by losses in our mirror coatings to $P_{out}^{max}=0.83 \pm 0.03$. Numerical simulations show that it should be possible to reach this value in our experiment \cite{maurer2004} (that is, the probability of the ion emitting into the cavity mode could be near 100\%) and recent experiments with our system show that $P_{out}\approx 0.7$ should be possible by cooling the ion close to the axial mode ground state (and thereby enhancing the coupling strength of the cavity-mediated Raman transition, in comparison to the detrimental spontaneous scattering rate). 

Finally, the achieved photon conversion stage efficiency is predominantly limited by unwanted excitation of higher-order spatial modes in the involved PPLN ridge waveguides \cite{Krutyanskiy2017}. A total device efficiency of 0.5 should be within reach with more careful attention to coupling into the guides and minimising other passive optical losses. Combing all of the aforementioned improvements would lead to a total 50~km detection probability of nearly 0.03, close to the fibre transmission of 0.1.  

Note that lower loss telecom fibres, than the one used here, are available (0.16 dB/km, Corning SMF-28 ULL) with a corresponding 50km transmission of 0.16 and any improvement in fibre technology will further increase that value.


\subsection{State characterisation}
\label{sec_State_characterization}
To reconstruct the ion-photon state, a full state tomography of the two-qubit system is performed.
On the photon polarisation qubit side, the state is projected to one of 6 states (horizontal, vertical, diagonal, anti-diagonal, right circular and left circular) by waveplates and polariser. 
This is equivalent to performing projective measurements in three bases described by the Pauli spin-1/2 operators. For example, horizontal and vertical are the eigenstates of the Pauli $\sigma_z$ operator. 
On the ion qubit side, measurement is performed in the three Pauli bases as described in section \ref{pulse_sequence}. 

For each of the 9 possible joint measurement bases (choice of photon basis and ion basis), the numbers of events corresponding to one of the four possible outcomes of these 2-qubit measurements are considered. 
We then divide the number of events recorded for each outcome by the total number of events recorded for the given basis (divide each number by the sum of four) and thus obtain estimates of the outcome probabilities.   
These probabilities are used to reconstruct the 2-qubit state density matrix by linear search with subsequent Maximum Likelihood method \cite{PLA99_likelihood}.
The values of fidelity, concurrence and other measures presented in the main text are calculated using reconstructed density matrices for each of the experiments.

For statistical analysis (determining error bars in quantities derived from the reconstructed density matrix), the Monte-Carlo approach was implemented \cite{efron1986_MC}. 
Briefly, we numerically generate M = 200 sets of 36 event numbers with Poissonian distribution and mean value equal to the experimental value for each of the 36 possible outcomes. 
From these simulated event numbers we derive simulated outcome probabilities, the same way as we do for the experimental counts.
Then we reconstruct M density matrices for this simulated data and for each one we calculate the quantities of interest (fidelity, concurrence). The error bars given in the main text represent one standard deviation in the widths of the distributions of these quantities over M simulated data sets.  

We quantify the state quality in terms of fidelity $F^m$ defined as $F^m = \left[\operatorname{Tr} \sqrt{\sqrt{\rho_{exp}} \rho_{max.ent.}\sqrt{\rho_{exp}}}\right]^2$, where $\rho_{exp}$ is the density matrix, reconstructed from the experiment data and $\rho_{max.ent.}$ is the density matrix of the nearest maximally-entangled pure state. This nearest state is found by exposing a perfect Bell state to single qubit unitary rotations and searching for a state providing the best fidelity with the experimentally obtained one.


\subsection{Imperfections in the entangled state}
\label{sec_imperfections}
Sources of infidelity in the experimentally-reconstructed ion-photon entangled state given in the main text are now analysed. 
As we will show, the  50~km ion-photon state infidelity can be accounted for (to within statistical uncertainty) by taking into account background detector counts and imperfections in the initial ion - 854 nm photon state output from the ion-trap.  

Three independent experiments are performed, corresponding to state tomography of the ion-photon state at three different points in the path. 
First the ion-854 nm photon state immediately at the cavity output (using free space polarisation analysis and two single-mode fibre-coupled 854~nm photon detectors, one at each port of a polarising beam splitter).
Second, the ion-1550 nm photon state immediately after conversion (with only a 1~m telecom fiber), referred to as $0 ~\rm{km}$ distance.
Third,  the ion-1550 nm photon state after 50~km travel (as presented in the main text).
The reconstructed state fidelities, with maximally entangled states, are presented in table \ref{infidelities} (bottom row `Experiment').

First the effect of background photon detector counts is analysed (defined as any reason that the relevant photon detector clicks other than a photon from the ion).  
For this, the background count rate is extracted from the measured counts in the tomography experiments by looking far outside the time window in which the ion-photon arrives (outside the time-delayed Raman pulse window), giving $2\pm 0.1~\rm{cps}$ for the 1550 nm photon at 50 km and  $10\pm 1~\rm{cps}$ for the 854 photon, which are both in agreement within the telecom (1.9$ \pm{0.15}$ cps) and 854~nm detectors' (10.1$ \pm{0.9}$ and 10.8$ \pm{1}$ cps) dark count rates respectively (measured independently). For the 1550~nm photon at 0 km we get $4\pm 0.1~\rm{cps}$, where the additional 2 cps background is produced by the photon conversion pump laser anti-Stokes Raman scattering which was reported in \cite{Krutyanskiy2017}. Note that this added noise is attenuated at the same rate as the photons from the ion over the 50~km, and so  becomes a small contribution to the background compared to the intrinsic detector dark counts (which do not attenuate over distance).  

The infidelity that the background counts will contribute when applied to a perfect maximally-entangled Bell state is simulated numerically. Specifically, the expected background count probability in our photon time-window is added to the expected measurement outcome probabilities for a perfect state, then a new `noisy' state density matrix is  reconstructed via Maximum Likelihood tomography. We call this approach `Model 1', which simulates the effect of measured background counts only, and find that it explains the majority of the infidelity in the 50~km state (see Table \ref{infidelities}).  

Model 2 takes, in addition to the background counts, the measured imperfect 854~nm ion-photon state into account. That is, the tomographically reconstructed ion-854nm-photon state is used as the state to which background counts are added as with Model 1. The results, shown in table \ref{infidelities}, show that background counts and imperfections in the initial 854~nm state explain the state infidelities to within statistical uncertainty. 

Regarding infidelities in the initial ion-854 nm photon state: the fidelity in this case is limited by the state purity ($Tr(\rho^2)=0.94\pm 0.01$, where  $\rho$ is the 854~nm  reconstructed state) meaning that only the imperfections leading to decoherence (or effective decoherence) need be considered. Possible error sources include errors in the 729~nm laser pulses used to determine the ion measurement basis, decoherence of the ion-qubit due to e.g.  fluctuating magnetic fields and relative intensity fluctuations of the two frequency components in the Raman drive leading to a mixture of different states over the duration of the experiment. Identifying the size and relative contribution of these errors is beyond the scope of this work. The achieved fidelity at 854~nm is similar to that achieved in \cite{Stute2012}.  

In conclusion, the fidelity of the 50~km ion-photon state is limited by detector dark counts.

\begin{table}[h]
\vspace{2mm}
\begin{tabular}{|c|c|c|c|}\hline
 Fidelity, \% & 854nm{@}0km & 1550{@}0km & 1550{@}50km  \\ \hline
 Model 1 & 99.5 & 96 &  86 \\ \hline
 Model 2 & - & 93 &  83 \\ \hline
 Experiment&$96.7\pm 0.6$   & $92\pm$ 2 &  $86\pm$3 \\ \hline
\end{tabular}
\label{infidelities}
\caption{Comparison of fidelity with a maximally-entangled state ($F^m$) for simulation of different noise models and experiment. Model 1: Bell state subjected to background counts during photon quibit measurement.  Model 2: Experimentally reconstructed 854~nm state affected by background counts.}
\end{table}


\subsection{Quantum Memory}
\label{sec_memory}


One of the functions played by quantum matter in a quantum network is as a memory to store established entanglement, while entanglement is being made or processed in other parts of the network. 
Decoherence processes in the matter qubit will limit the distance over which it is possible to distribute quantum entanglement (the distance a photon could possibly travel in the `coherence time' of the matter qubit). 
%
In the 50~km experiment in the main text, the ion qubit is already stored for the 250~$\mu$s photon travel time over the 50~km fibre, with no statistically significant reduction in the ion-photon entanglement quality \footnote{This was achieved by installing a mu-metal shield around the ion-trap vacuum to attenuate ambient magnetic field fluctuations.}. 

Additional measurements are performed to see how long ion-photon entanglement could be stored in our current ion-trap system. 
Specifically, state tomography is performed for increasing delays introduced between detecting the telecom photon (0~km fiber travel distance) and measuring the state of the ion-qubit.
This is equivalent to introducing an additional storage time for the ion-qubit. 
The results show that strong entanglement is still present after 20~ms wait time ($F^{m}=0.77\pm 0.04$, $C=0.57\pm 0.08$), the longest wait time we tried. 
This already opens up the possibility of distributing entanglement over several thousands of kilometers and the time to perform hundreds of single and multi-qubit ion quantum logic gates \cite{Lanyon57}. 
Dominant sources of decoherence of our ion-qubit are uncontrolled fluctuating energy level shifts due to intensity fluctuations of the 806~nm laser field used to lock the cavity around the ion and fluctuations in the local ambient magnetic field due to a nearby elevator. 
Further attention to minimising the absolute size of these fluctuations should lead to entanglement storage times of $\approx$ 100~ms and therefore the possibility to distribute entanglement to the other side of the earth. 
Beyond this, the ion-qubit could be transferred to hyperfine clock transitions within different co-trapped ion species that offer coherence times of many seconds and longer \cite{Wang2017}. 

\end{document}